# Imaging the inside of thick structures using cosmic rays

E. Guardincerri[a,1], J. M. Durham[a], C. Morris[a], J.D. Bacon[a], T. M. Daughton[a], S. Fellows[a], O. R. Johnson[a], D. J. Morley[a], K. Plaud-Ramos[a], D.C. Poulson[a], Z. Wang[a]

a) Los Alamos National Laboratory, Los Alamos, NM, 87545 USA

**Abstract:** The authors present here a new method to image reinforcement elements inside thick structures and the results of a demonstration measurement performed on a mock-up wall built at Los Alamos National Laboratory. The method, referred to as "multiple scattering muon radiography", relies on the use of cosmic-ray muons as probes.

The work described in this article was performed to prove the viability of the technique as a means to image the interior of the dome of Florence Cathedral Santa Maria del Fiore, one of the UNESCO World Heritage sites and among the highest profile buildings in existence. Its result shows the effectiveness of the technique as a tool to radiograph thick structures and image denser object inside them.

## Introduction

The cathedral of Santa Maria del Fiore (Saint Mary of the Flower) was founded in 1296 and is the main church of Florence. The church of Florence is one of the UNESCO World Heritage sites. Its majestic dome ("Cupola" in Italian) was constructed between 1420 and 1436 under the direction of Filippo Brunelleschi and now dominates the Florence skyline.

---

[1] Corresponding author: elenaguardincerri@lanl.gov

The Cupola consists of two shells, with the inner shell 2.25 m and the outer shell 0.8 cm thick at the base, separated by a 1.2 m and connected by spurs or buttresses. It is built out of large blocks of sandstone up to an elevation of about 7 m above its base. From here to the base of the lantern, the masonry is made of clay bricks of different sizes and arranged according to different patterns with the purpose of ensuring the Cupola's structural stability. At the time of its construction, the Cupola was the largest dome ever built and it is still not clear how Brunelleschi managed to build it without a temporary supporting structure (centering). Three pairs of *macigno* (large stones) chains are believed to reinforce the Cupola, although only the existence of the lowest pair has been proven. A wooden chain was also installed 7.75 m above the base of the Cupola. The base of the dome is octagonal in shape, the apothem of the polygon describing its inner base surface measures 20.46 m [1]. According to some scholars, there are iron chains inside the masonry, although investigations with metal detectors have failed to provide conclusive evidence. What is certain is that thousands of Florentine pounds of iron (1 Florentine pound = 0.34 kg) were purchased during the construction of the Dome though only a small fraction of it is currently accounted for.

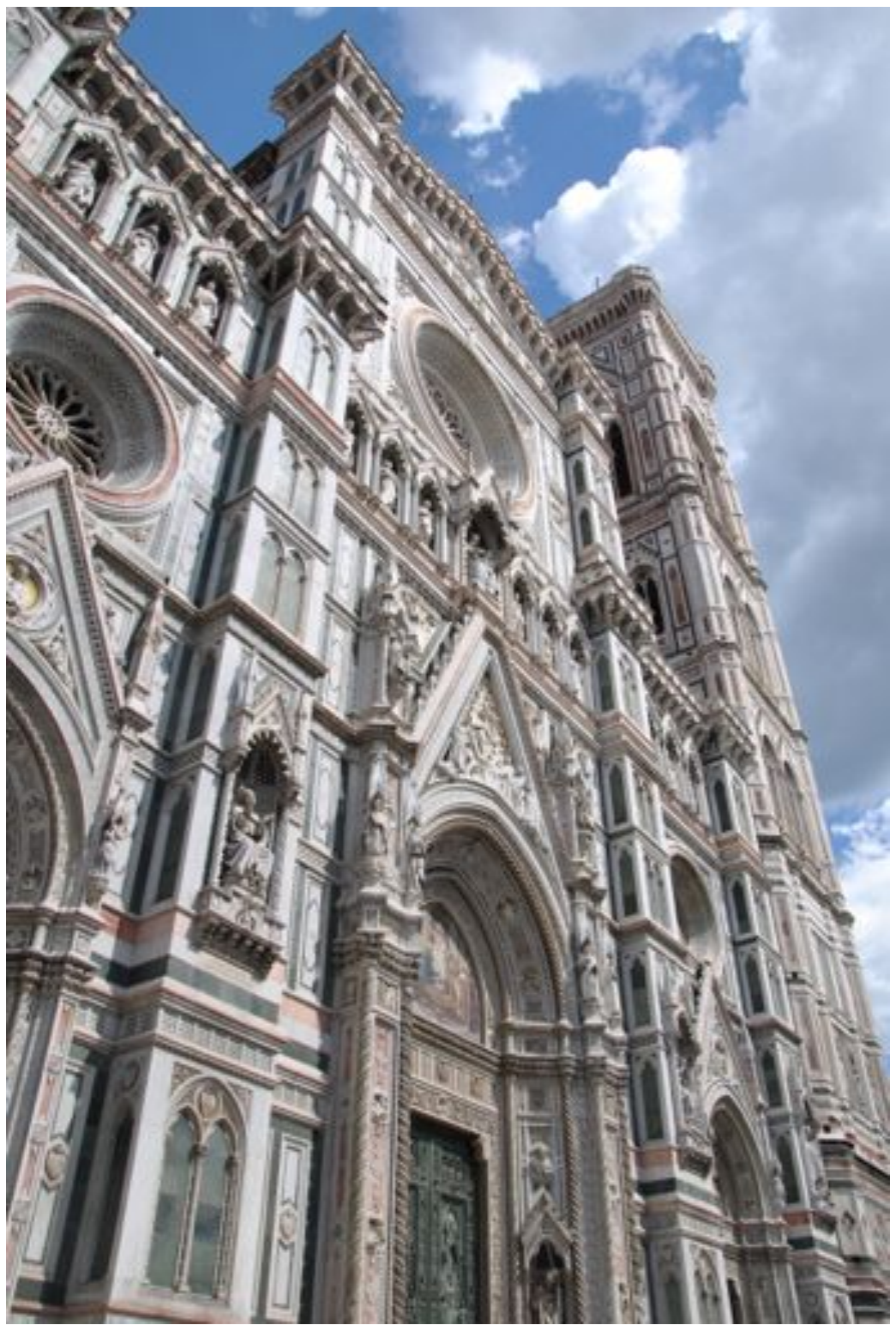 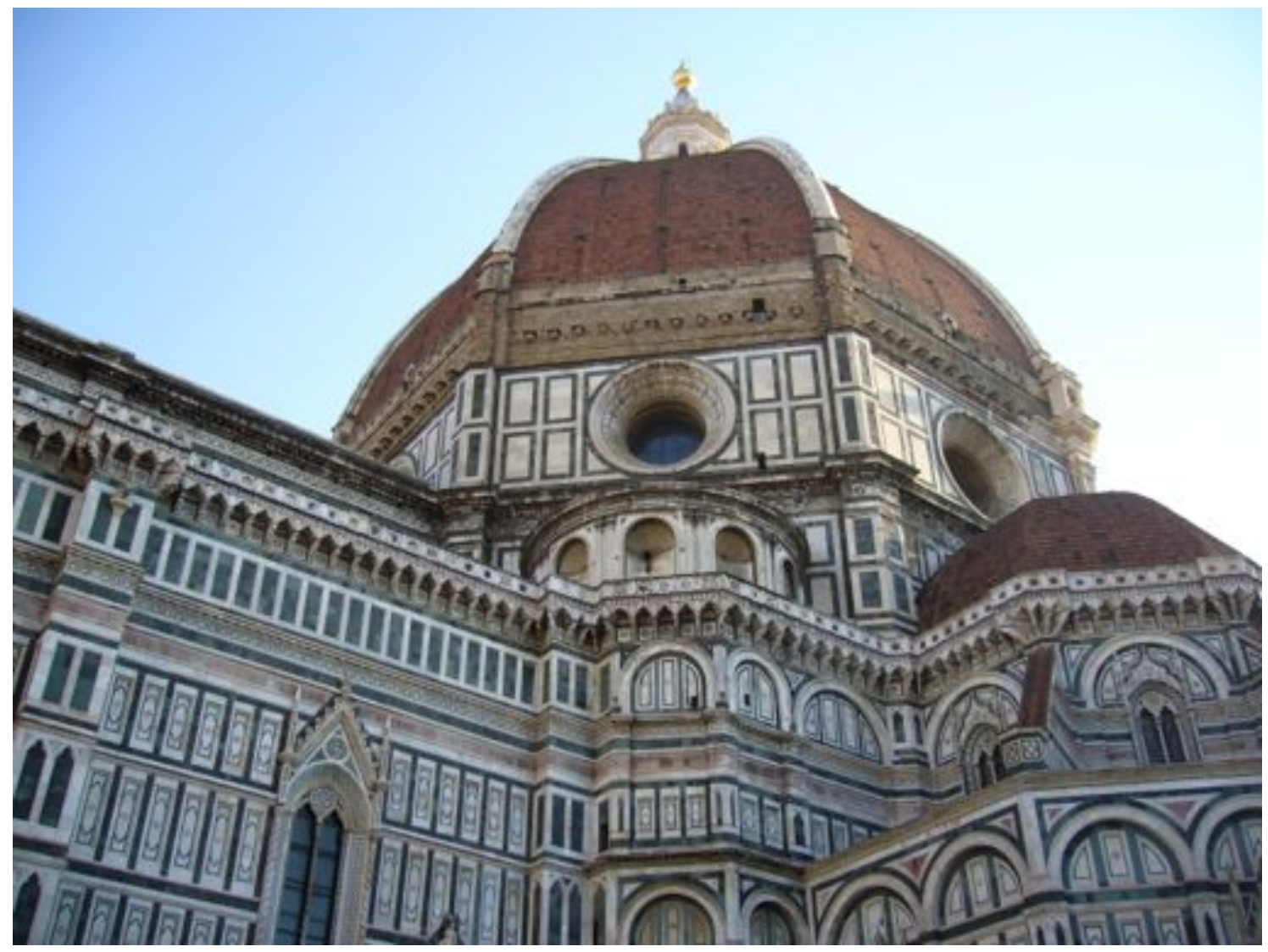

**Figure 1:** Pictures of the Cathedral of Santa Maria del Fiore, images by authors. Left insert: the facade. Right insert: part of the main nave with Brunelleschi's cupola above the transept.

The Dome has been affected by cracks for centuries; some of them are currently 6-8 cm wide. It is likely that they appeared shortly after the completion of the Cupola, although they were first mentioned in 1639. The location of the main cracks is shown in Figure 2 and Figure 3. The width of the cracks increases at a rate of 7.5 mm/century [1].

Detailed knowledge of the structure of the Cupola would benefit the finite element calculation models that are used to evaluate its behavior under static and dynamic (earthquake) conditions. Only the parts of the monument that are directly visible have been extensively studied because the techniques capable of surveilling the interior structure are limited. Brunelleschi was very secretive about his work and purposely did not leave technical drawings.

The authors realized that multiple scattering muon radiography could be used to image the inside of the Dome's walls. Muons are charged leptons with a mass of 105.6 MeV/$c^2$, about 200 times the mass of an electron. Cosmic-ray muons are naturally produced in the atmosphere by the interactions of primary cosmic rays, mostly protons, with the nuclei present in the atmosphere itself. They reach the surface of the Earth at a rate of about 1/$cm^2$/min with a mean energy of $\sim$4 GeV [2]. They have no hadronic interaction with nucleons, and their relatively large mass limits energy loss due to bremsstrahlung radiation. These properties allow energetic muons to penetrate large amounts of material that are inaccessible to other particles. Muons do possess an electric charge and undergo Coulomb scattering off nuclei as they pass through matter.

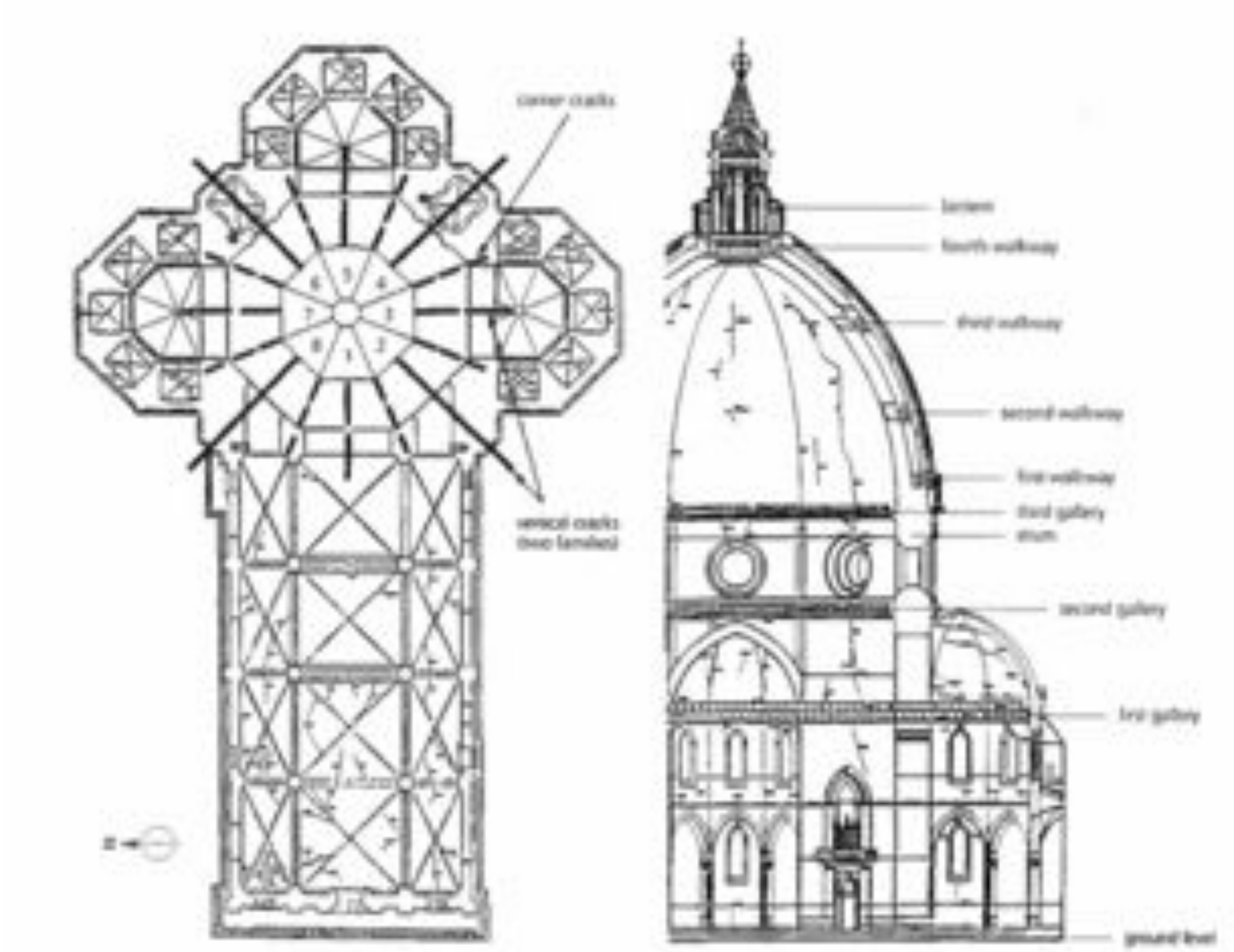

**Figure 2:** plan (left) and cross-section (right) of Santa Maria del Fiore showing the location of the main cracks.

Reprinted from reference [1] with permission.

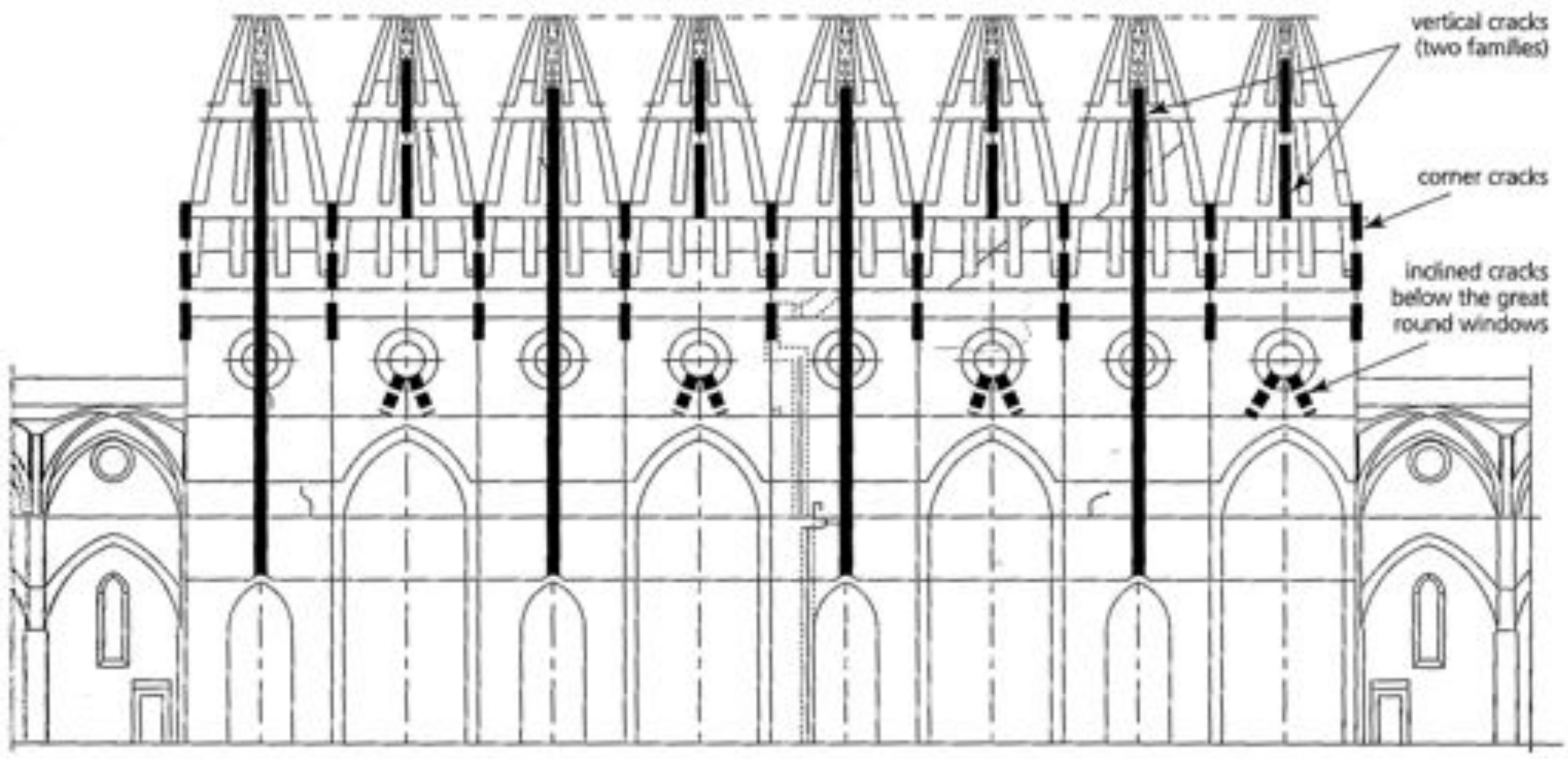

**Figure 3:** Two dimensional representation of the dome showing the main cracks. Reprinted from reference [1] with permission.

The first use of cosmic ray muons for radiography was in 1955, when George measured muon attenuation to determine the overburden of rock above a tunnel [3]. This was followed by Alvarez et al., who used this method to confirm the Second Pyramid of Giza did not contain any undiscovered chambers [4], and more recently by several groups examining geologic features [5][6][7][8][9].

A different method, developed at Los Alamos National Laboratory, uses measurements of the multiple scattering angle of individual muons passing through an object to create tomographic images of the object's interior structure [10]. This technique was originally developed to inspect cargo containers for illicit trafficking of nuclear material [11][12], and has since been applied to detection of Special Nuclear Materials [13], industrial corrosion [14], nuclear reactors [15][16][17], and will be used to determine the condition of the damaged cores of the Fukushima Daiichi nuclear power plant [18][19].

The method is based on the fact that muons, when travelling through materials undergo many single Coulomb scatterings with the charged atomic nuclei. As a result, they are deviated from their initial trajectory and exit the material with a different direction from the original one.

The theory of multiple coulomb scattering was developed by Moliere and Bethe [20][21][22]. It predicts, for the muons that go through an object, a Gaussian polar angle distribution with tails. The dominant part of such distribution can be described by:

$$\frac{dN}{d\theta} = \frac{1}{2\pi\sigma^2} e^{-\frac{\theta^2}{2\theta_0^2}} \qquad (1)$$

where $\theta$ is the polar angle and $\theta_0$ is given by

$$\theta_0 = \frac{14.1MeV}{p\beta} \sqrt{\frac{l}{X_0}} \qquad (2)$$

with $p$ being the muon momentum, $\beta$ its velocity, and $X_0$ the radiation length for the material. By measuring the distribution of cosmic muons scattering angles after they traverse an object of interest, an image of the object itself can be reconstructed as described in the paper by Borozdin et al. [18]. Since muons naturally and constantly shower the Earth, no additional radiation source is needed, making this technique completely safe; this avoids the complications related to the handling of artificial radiation sources or particle accelerators.

By using multiple scattering muon radiography on the inner wall of the Florence Dome, it is possible to:

- Identify and image iron elements inside the masonry of the dome.
- Determine a crack's profile inside the wall, sheding light on how the wall itself is built. In fact, cracks propagate along weakness in the masonry, normally along the mortar as compared to the bricks. This information in particular could tell whether the inner wall is entirely made of bricks or if rubble masonry was used.
- Image the transition between the stone and the brick masonry at the base of the Dome.

In order to prove the feasibility of this measurement, a demonstration measurement was performed at LANL during the summer of 2015.

**The measurement**

If any reinforcement structure is present in the masonry of the Cupola's wall, it most likely has to be located inside the inner shell. This shell is much thicker than the outer shell, and thus plays a more important role in sustaining the whole structure. A concrete wall having the same thickness, in radiation lengths, as the inner wall of the Dome was built. The masonry of the Dome is mostly made of clay bricks. In order to calculate their radiation length a chemical composition and a density indicative of clay bricks from the same historical period and geographical area was obtained from an analysis of ancient bricks by Fernandes et al. [23]. The bricks were assumed to have the following chemical composition, the percentages being weight fractions: 55% $SiO_2$, 25% $Al_2O_3$, 10% $FeO_2$, 5% $K_2O$, 5% $H_2O$. The density was chosen to be 1.6 $g/cm^3$, a typical value for the clay bricks manufactured in the same period of time and geographical area. The mock-up wall was built from standard radiation shielding concrete bricks whose radiation length was 26.7 g/cm$^2$ based on the estimates provided in the Review of Particle Physics [2] and whose density was measured to be (2.13±0.13) g/cm$^3$. From these values, the thickness of a concrete wall equivalent to 2.25 m of clay was calculated to be (1.74 ±0.01) m. Since the size of the available bricks was not an integer fraction of 1.74 m, the thickness of the actual wall was actually (1.73±0.01) m.

Three iron bars were placed inside this mock-up wall, in a dedicated empty space purposely left in the middle of it. The mock-up wall is shown in Figure 4, the iron bars are shown in Figure 5.

The bars had rectangular or square cross-sections with the following dimensions:

- 4.76 cm x 5 cm.
- 2 cm x 3 cm, the bars in the Cupola wall are believed to be this size [24].
- 10 cm x 10 cm.

The LANL Mini Muon Tracker (MMT) [25], was used to perform the demonstration measurement. The detector consists of two modules made of 576 drift tubes arranged in planes. Each of the two modules consists of 6 planes of drift tubes, three of which are oriented along the horizontal X direction, and three of which are oriented along the horizontal Y direction, perpendicular to X. Each module can independently track cosmic rays and the trajectory information can be used to generate images of the object located between them.

The two modules of the MMT were deployed on opposite sides of the wall as shown in Figure 6.

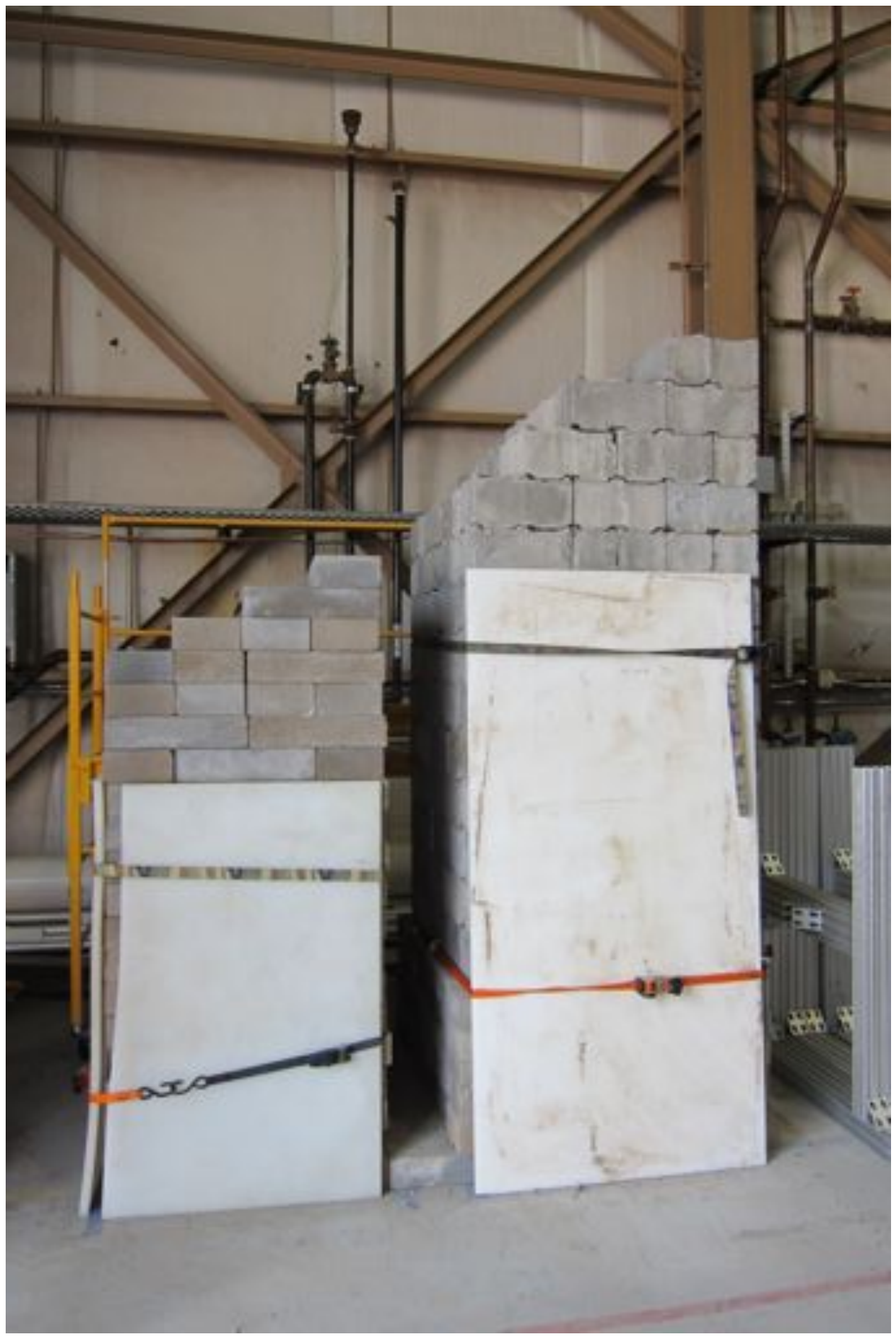

**Figure 4:** The mock-up wall after its completion. Image by authors.

The rate of cosmic-ray muons increases as their direction approaches the vertical. For this reason, one of the two tracking detectors was raised by 121.92 cm (4 ft) to increase the counting rate and reduce the data taking time. We collected data for 35 days, during which we collected a total of 3,112,277 muon tracks through both detectors.

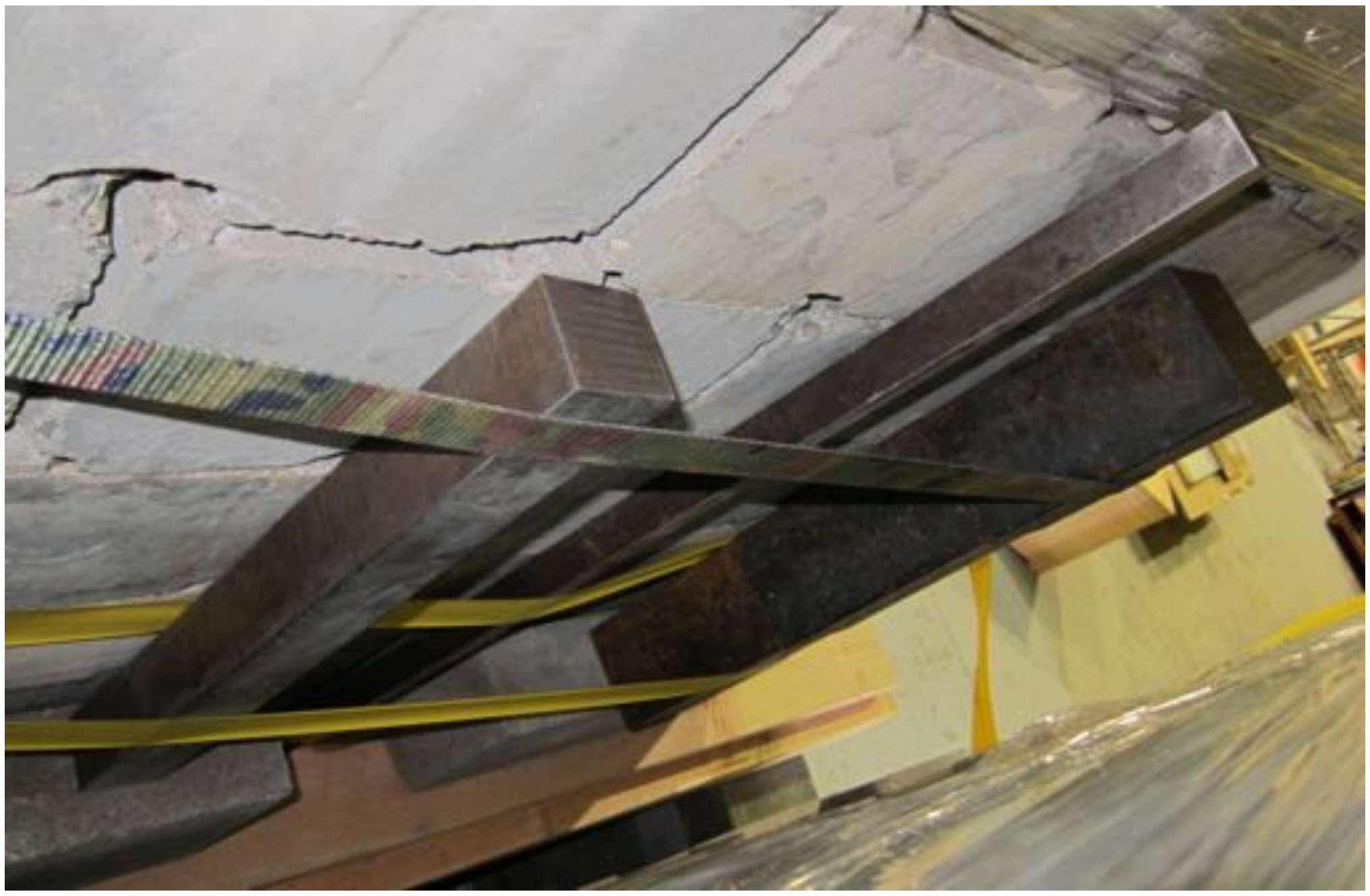

**Figure 5:** The iron bars inside the wall. Image by authors.

An image of the vertical plane passing through the iron bars was reconstructed using the algorithm described in the paper by Morris et al. [26] and the resulting image is shown in the left insert in Figure 7.

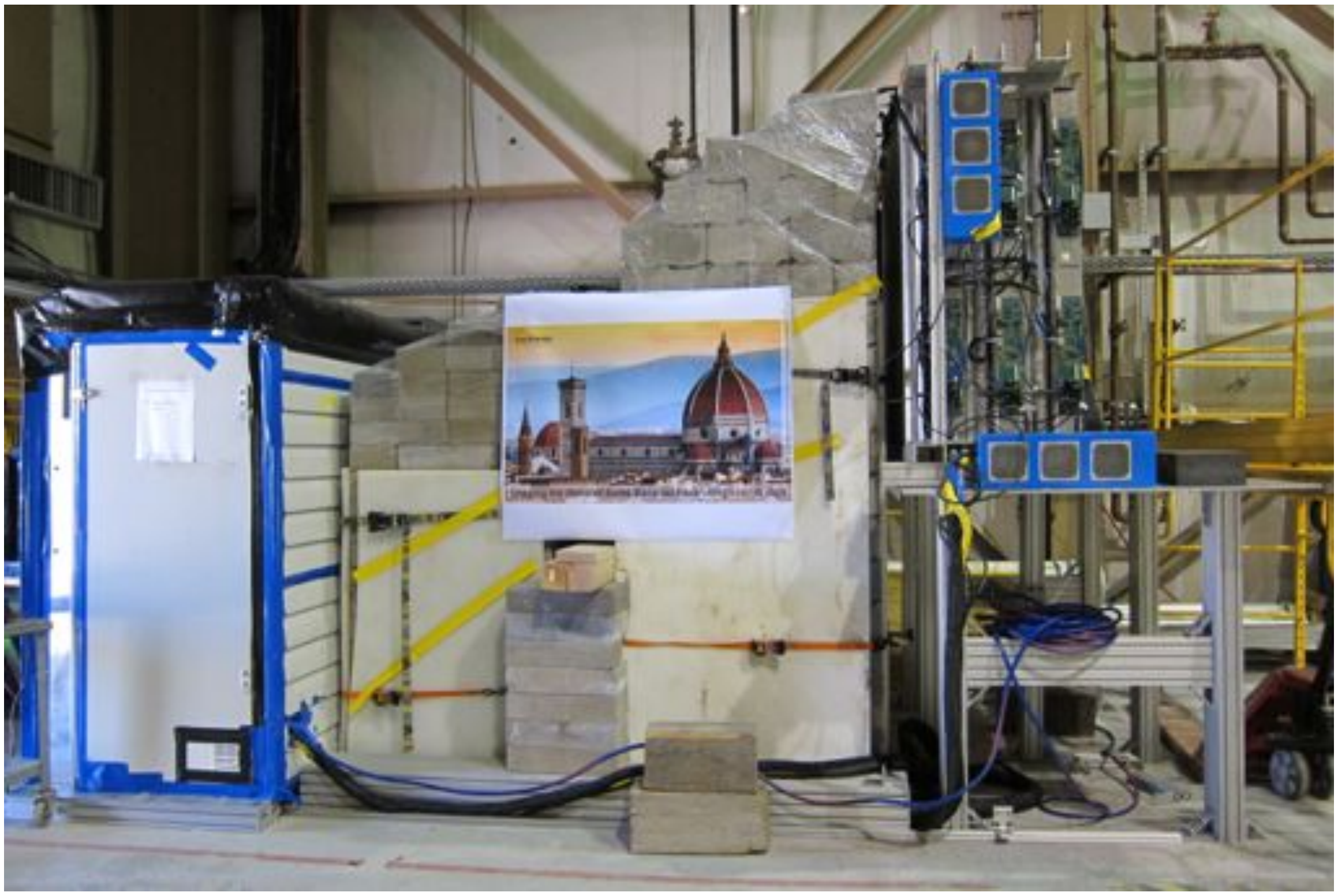

Figure 6: The Mini Muon Tracker taking data around the mock up of the Cupola wall. The lower detector is still enclosed in a shed after a previous measurement campaign during which it needed to ne weather-proof. Image by authors.

The pixels in the image have 1cm x 1cm size and their intensity is proportional to the areal density parameter extracted from the fit [10].

All the iron bars are visible, including the thinnest one. For comparison, a Monte Carlo simulation of the same experimental set-up was performed with GEANT4 [27] and its data was analyzed with the same analysis technique. The number of simulated tracks going through both detectors used in the analysis was the same as the number of through going tracks collected in the real experiment. The areal density image obtained from the Monte Carlo simulation is also shown in Figure 7. Both the three bars and the concrete wall are uniform and homogeneous along the vertical direction. The 2D images shown in Figure 7 were therefore integrated along the vertical direction, and the 1D plot shown in Figure 8 was obtained.

In order to better reproduce the experimental data, the Monte Carlo image had to be scaled by a factor of 1/1.29. The authors believe this to be due to differences between the muon momentum spectrum used in the simulation and the actual

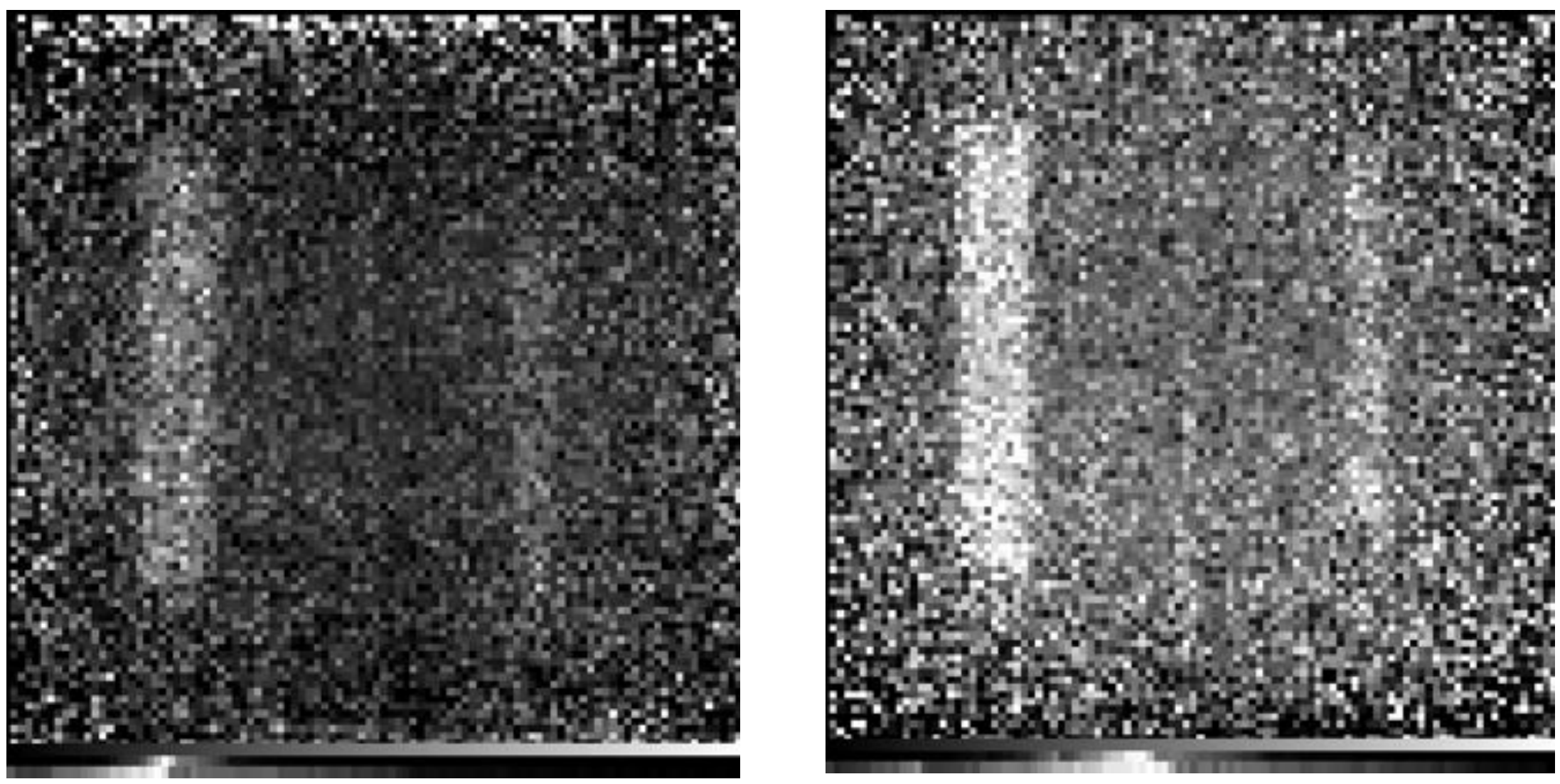

**Figure 7:** Left: Image of the three iron bars described in the text inside the concrete mock-up wall, obtained using 35 days of data collected with the MMT. Right: Image of the same three bars obtained from the output of a Monte Carlo simulation.

muon momentum at the measurement site, as already noticed in one of their previous works [28].

The contrast and the resolution seen in the two images are very similar.  A slight difference between the two histograms in Figure 8 can be noticed at the extremes of the horizontal scale (x < 360 mm and x > 960 mm). This is probably due to bias in the experimental data in proximity with the edges of the detector's field of view.

The good agreement between the data and the simulation demonstrates that the technique is well understood and that the performances of future measurements can be confidently predicted.

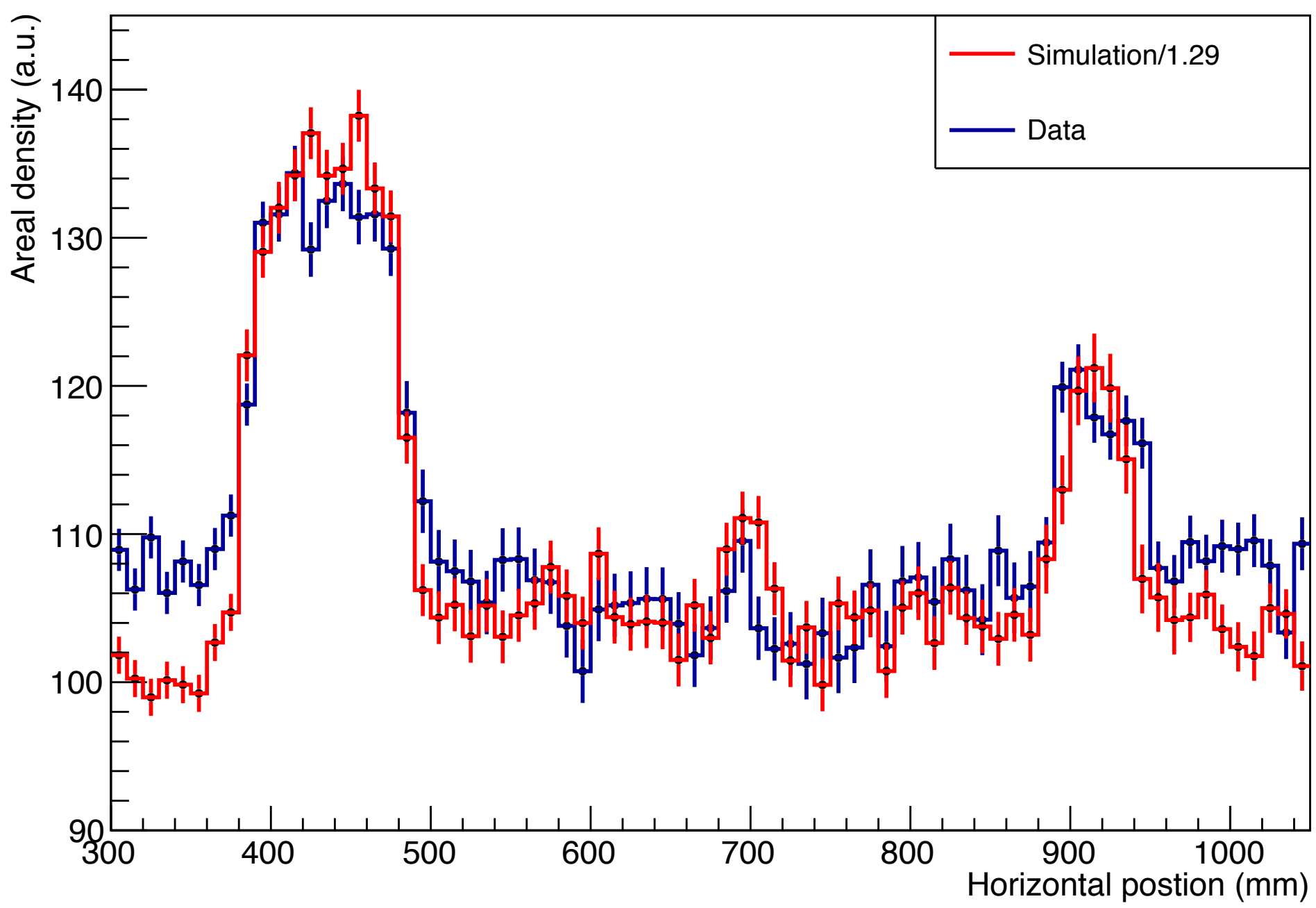

**Figure 8:** The integral of the areal density obtained from the data (blue) and from the Monte Carlo simulation described in the text (red).

The demonstration measurement discussed here was performed in Los Alamos, NM, USA, at an altitude of about 2,230 m above sea level, to be compared with the 50 m elevation of Florence, Italy.

Since the near horizontal muon flux at sea level is about 5 % lower [2] than it is at ~2,200 m elevation a measurement at sea level, in general, would take 5% longer than the same measurement performed at 2,200 m altitude.

The authors focused here on the problem of locating iron reinforcement structures inside a thick wall. Showing the potential of muon radiography to image the profile of a crack inside a thick wall requires a different mock-up wall, with a 6-8 cm wide crack in it. Multiple measurements need to be taken at different angles around the crack to obtain a 3D tomographic image, and further development on software to reconstruct the actual tomographic image is also necessary. The authors are currently working on this problem, which will be the subject of a dedicated publication.

**Conclusion**

The authors have shown that multiple scattering muon radiography can effectively be used to locate and image iron bars inside of a 173 cm thick concrete wall, making the method ideal for investigating the interior of Brunelleschi's dome.

This method can, in general, be used to radiograph thick structures, imaging denser objects inside them (reinforcement bars being an outstanding example).

Applications include, but are not limited to, the monitoring of building and civil constructions (e.g. dams, bridges) to assess their structural stability and the imaging of the interior of thick archaeological artifacts.

**Acknowledgements**

This work was funded by the laboratory Directed Research and Development program at Los Alamos National Laboratory.

The authors would like to thank Prof. Carlo Blasi, Dr. Michele Fanelli and Dr. Federica Ottoni for the useful discussion and advice.